\documentclass[journal]{IEEEtran}

\ifCLASSINFOpdf
  \usepackage{graphicx}
  \usepackage[colorlinks=true,linkcolor=blue,citecolor=blue,pdfmenubar=true]{hyperref}
  \usepackage{rotating}
  \usepackage{booktabs}
  \usepackage{multirow}
\else
\fi
\begin{document}
%
\title{Development of an Android Application for an Electronic Medical Record System in an Outpatient Environment for Healthcare in  Fiji}
%
%

\author{Daryl Abel, Bulou Gavidi, Nicholas Rollings
        and Rohitash Chandra
\thanks{Daryl Abel, Bulou Gavidi, Nicholas Rollings and Rohitash Chandra are with the School of Computing, Information and Mathematical Sciences, Faculty of Science, Technology and
Environment, University of the South Pacific, Suva, Fiji. They are also with Artificial Intelligence and Cybernetics Research Group, Software Foundation, Nausori, Fiji (aicrg.softwarefoundationfiji.org) (Email: c.rohitash@gmail.com)}
}

%
%

\markboth{Technical Report, AICRG, Software Foundation, Fiji, March 2015}{}%
%



\maketitle

\begin{abstract} 

The outpatients department in a  developing country  is typically understaffed and   inadequately equipped to handle a large numbers of patients filing through on an average day.  The use of electronic medical record (EMR) systems can resolve some of the  longstanding medical inefficiencies common in developing countries.  This paper presents the design and implementation of a proposed outpatient management system that enables efficient management of a patient's medical details.  We present a system   to create appointments with medical practitioners by integrating a proposed  Android-based mobile  application with a selected open source EMR system. The application allows both the patient and the medical practitioners  to manage appointments and make use of the electronic messaging  facility to send reminders when the appointed time is approaching in real-time.  A mobile application prototype is developed and the  road map for  implementation is also discussed.

\end{abstract}

%
\IEEEpeerreviewmaketitle

\section{Introduction}

An Electronic Medical Record (EMR) is a digital version of a patient's paper-based medical record [1]. Its usage allows for accurate and detailed patient health record-keeping and results in faster and efficient search and retrieval.

The use of Electronic Medical Record (EMR) technology has already begun [2]. Although the pace is  slow in developing countries, it is evident that more health care organizations are moving from a paper-based system to EMR systems in developed countries [3]. 

It has been reported that the United State's Health Care industry is 20 years behind the rest of the nation’s industries [4]. The adoption of EMR systems have taken place in both developing and developed countries however, developing countries in particular must overcome enormous challenges that include but are not limited to a lack of ICT infrastructure and the necessary technical support to maintain such health care systems [5].


 OpenEMR  is a free open source medical software system designed particularly for developing countries that do not have the budget to use proper healthcare management systems. The OpenEMR is successfully deployed in healthcare clinics in many countries including Canada, Kenya and the Bahamas [6]. 

Mobile phones are a candidate platform for delivering and retrieving health information due to its widespread adoption, and technical capabilities [8]. Its functionalities may include retrieving a patient's medical records any time and anywhere, and the generation of tailored messages to patient's smart phones as a reminder of his/her appointment a few hours before the appointed time.


Health care centres in Fiji  use a manual paper-based filing system seemed to be the norm rather than the exception.  Patient Information System (PATIS) is the major software used by Fijian hospitals \cite{Freeman2010},  but not available in health centres that serve a major segment  of the population \cite{FijiMOH2014}. 
  There have been significant enhancements to Fiji's PATIS, which now has functionality usually only provided by specialised clinical system modules \cite{Soar2012}. A review team in 2010 was informed that the PATIS system is being transferred to one that is web based and this will help encourage wider use of the very valuable data being generated by the system \cite{Freeman2010}. The latest version of PATIS with improved accessibility and speed is referred to as PATIS-Plus \cite{FijiMOH2014}.

OpenEMR is an open source software used in health information systems \cite{Feufel2011e85} \cite{Ford2006106}. Open source software solutions are becoming popular in developing countries as they can be deployed taking into consideration the cost of purchasing licensed software \cite{Kiah2014} \cite{WikipediaOpenSource}.


The typical outpatient environment, as experienced in Fiji, is chaotic and time consuming [9] where patients must wait in a queue without any indication of when they will be seen by the doctor. This is a situation commonly found in outpatient clinics and walk-in appointments. The waiting times can be greatly reduced if a patient is allowed to make an appointment using a smart phone and to arrive at the hospital at the appointed hour. Also, a doctor is usually unaware of the size of the patient queue, unless he checks with a nurse or enters the waiting room thus our proposed system also integrates a doctor's app that shows his current appointments.

The absence of an appointment system results in long waiting times for patients in health centres and hospitals. A 4 to  8 hour awaiting time is common. Currently, patients do not have instant access to their medical records but the surge of mobile phone usage in Fiji suggests a possible solution to this problem.

We propose an system for Fiji's healthcare industry to take advantage of mobile devices and use mobile software  applications. The proposed system will only focus on improving the doctor-patient appointment process within an outpatient environment. The proposed system will allow  health care centre administrators and the doctor to efficiently manage patient information and appointment details which is centrally stored in a cloud server. This should result in improved services to patients, improved operational efficiency by minimising errors in data entry, and an increase in revenue for the healthcare centre as more patients can be served. It has been found that such interactive systems, as opposed to one that do not provide health data for patients, will attract many users [10].

OpenEMR  is used to demonstrate the issue with compatibility of the proposed Android application. The goal is to configure the connectivity of the established databases for patent information and access them with the proposed  Android application in order to implement mobile  based appointment system. 

The rest of the paper is organised as follows. An overview of related work on EMR and applications  in Section II. Section III gives the details of the proposed  the design of the system architecture of the mobile application and Section IV gives the implementation of a prototype  with results on software testing.    Section V concludes the paper with discussion of further research. 

\section{Related Work}

An electronic medical record system  allows for the management and  collection of health information for  patients   which  also impacts in terms of improving work flows [11]. Natural disasters such as Hurricane Katrina in the United States underlined the reality of inefficiently stored medical paper records of scores of patients [12]. In addition, only very recently, the Obama administration implemented electronic health care programs nationwide in the \$2.4 trillion health industry [13]. A study carried out in 2011 indicated that 57\% of office physicians in the U.S use EMR [14]. Though this figure might not be representative of other countries, it gives an idea of how much attention is currently given to the use of EMR. 

EMR efficiently manages patient records [12, 15-17], minimizes error entries (consistent format is maintained) and generates comprehensive reports on patient histories and services [15]. It also facilitates instant communication between several actors as well as verifying details about medications [18] which in turn improves the quality of patient care [16, 19] and safety [2] as required information is accessed at the right time. Moreover, a few reports have indicated positive returns on investment as the doctor is able to serve more patients than it normally does [9, 20].

However, certain barriers may halt the adoption to the EMR system. For instance, physicians may not be that keen in transitioning to EMR system due to the complexity of maintaining EMR systems which increases financial cost [21]. Another author suggests that uncertainty about future mandates in regards to the EMR usage may also slow the implementation of such system [22].

Most of the reports reviewed tend to focus more on implementation, and less on design [23] which is a crucial factor in determining how readily the system would be accepted or rejected. Successful adoption is the result of achieving the right balance in aligning system functionality with requirements and working patterns of the target organisation [24].

Other barriers include the lack of capital requirement and financial support from the government, not finding the right EMR system to meet organization needs [2] and having low computer literacy [16] which is mostly found in the old age group.

Some may argue that allowing certain tasks which involves accessing a patient's personal details by practitioners within the EMR system may border on ethical issues such as data privacy, and reliability may be at stake. For this reason, some patients might think that it is necessary to limit a practitioner's access to patient information. However this could result in mis-prescribing medications. Therefore a few authors had proposed the development of a structured ethics framework to prevent unethical issues which may arise [25].

In addition, the use of EMR systems may have negative impacts especially on aged doctors who have become solely dependent on their manual charts where references for patients could easily be found by flipping pages and comparing details [26]. The new system may put a burden on these doctors as they will have to remember multiple patients' data. But a study that focused on EMR implementation in a large ophthalmology hospital in India [27] reveals that success came down to ensuring that the user interface was as close as possible to working with paper records, for instance, incorporating a drawing tool based on electronic pen-pad technology.

To overcome any initial resistance, or in worse case scenarios, a rejection of the EMR system at its introductory phase, the transition from a paper based one should clearly articulate the potential benefits and the degree of control that lies in the hands of physicians [25] and other clinical staff at the earliest stages of system development. Equally significant is the creation of a system that is operationally sustainable and does not rely on outsiders for support beyond the initial start up phase [26]. Recognising the need for a training program for doctor-patient communication including nurses [18, 30] is also necessary for training the system users to be ready in accepting the transition to the new system. Lastly, research and identifying vital factors during design and implementation will ensure that users will be able to use the system without difficulties [31]. 

As the development of patient-oriented health care services is on the rise, it becomes the responsibility of developers to ensure that the mobile application technology appropriately services the needs of the health intervention. Such health interventions must be simple to adjust to and alleviate the burdens associated with time and effort. Positive reactions from the community is dependent upon the minimal disruption to daily patterns of the average patient. Thailand, South Africa and China are among the few places where successful applications that sent daily reminders to patients are utilised thus showing an increased in medication adherence. [32]

\section{Proposed  System Architecture}

This section introduces the system architecture that underlies all processes involved in managing appointments for an outpatient environment. We propose this architecture for Fiji which could be later be implemented in  other Pacific Island countries.

\begin{figure*}
\centering
\includegraphics[height=10cm]{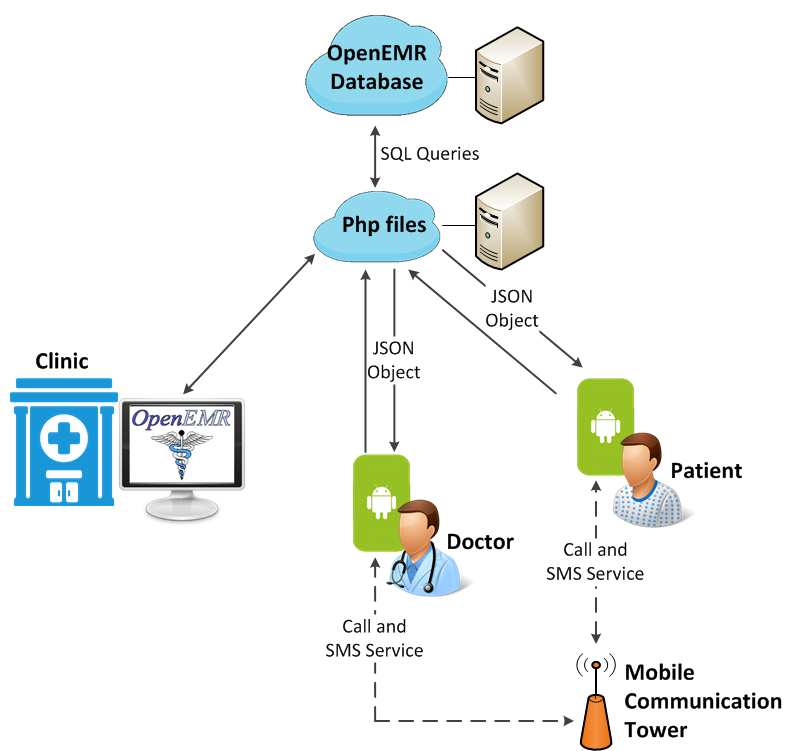}
\caption{Architecture of the proposed system. } 
\label{fig:1}
\end{figure*}


Patient Information Systems (PATIS) is not available in healthcare centres that heavily dominate outpatient services in Fiji. There is a strong dependency on the manual filing system for storing patient records. This becomes inefficient when patients are waiting in an outpatient queue to be served. Therefore, we proposed a system with the aim of minimising this problem. Since  PATIS is not a web based system, we use OpenEMR to demonstrate the implementation of the proposed framework where Android application use the databases from the main software. 

The proposed system architecture allows a patient to manage their appointments using an Android application as shown in Figure \ref{fig:1}). (Give some more sentences that gives more detail of the information flow, software etc in the architecture) 

The doctor uses another Android application to view his appointment list, view patient details and trigger an appointment reminder to the patient who will be seen in the next 30 minutes. The administrator primarily manages doctors' schedules as well as doctor and patient details (this includes registering a new doctor or patient) using the OpenEMR on a standalone computer within the clinic. All of which share data that is deployed on a MySQL database Server. This server provides the ability to store, retrieve and update patient, doctor and appointments details.

\begin{figure}
\centering
\includegraphics[height=7cm]{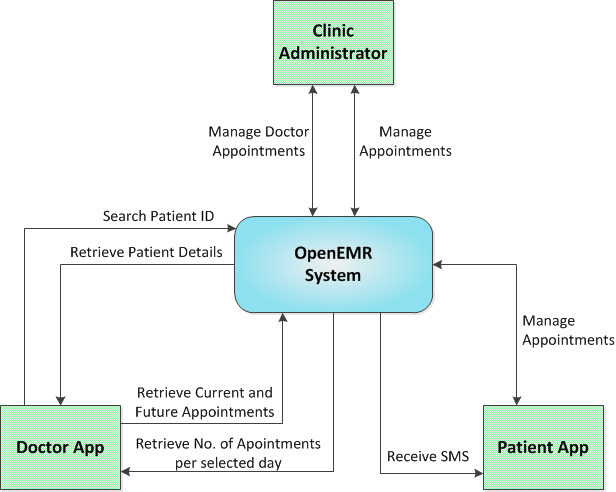}
\caption{Context diagram for the  proposed system. (Give some more sentences that describes the system in detail)}
\label{fig:2}
\end{figure}

The context diagram in Figure \ref{fig:2} illustrates the flow of information between the Android application  (App) and the OpenEMR system.(Give some more sentences that describes the system in detail. You need to fully describe the overall processes, information flow and so on)

The assumptions for this prototype were derived after observing the operations of a small clinic we visited. These assumptions are:-
\begin{enumerate} \itemsep0em
\item A morning (8am - 12pm) and afternoon (1pm - 5pm) shift
\item One doctor per shift
\item One hour lunch break
\item Each consultation lasts 15 minutes
\item Appointments can be made up to 30 days in advance but not more
\end{enumerate}

The following will further describe the functionalities of the prototype system developed:

\subsection{OpenEMR}

The administrator uses the openEMR to register/modify a patient or doctor details. All appointments added by patients are shown on the OpenEMR web application and in special circumstances, administrators can act on a patient's behalf in altering the patient's appointment upon the patient's request. After a patient is seen by the doctor, his/her appointment is removed from the OpenEMR calendar but still remains in the patient's history.

\subsection{Doctor's App}

The login screen (as well as a password resetting screen) is designed to authenticate doctors before accessing his Appointment list. After Logging in, a doctor may either search for a Patient using an identification number, view today's Appointments, view future Appointments or call the Pharmacist concerning medications. If there are existing Appointments for today, the screen will automatically display an alert dialogue with information of the first patient to be seen. This alert dialogue gives options for the doctor to either view more details of that patient or remove that patient's Appointment from the Appointment list after his/her appointment. A doctor can also send SMS reminders to all patients at the start of the day however an SMS reminder is always automatically send to the patient whose appointment is 30 minutes from the current appointment.

\subsection{Patient's App}

The Patient App has similar features like the {\it Login, Password Resetting} and {\it Viewing Personal Details} functions of the Doctor App. However the patient will not be able to alter their personal details via the app as this should only be the administrator's work. 

On the {\it Appointments} window, an alert dialogue containing information of the next appointment will pop up, giving the option to edit or delete the selected appointment. 

Adding an appointment allows a patient to select a date from a calendar. After the desired date is chosen, the available time slots are automatically shown in a list view to choose from along with the doctors for the morning and afternoon shift for that selected day. After adding an appointment, an appointment ID will be generated. This appointment ID will be sent by SMS to the patient 30 minutes before his/her appointment.
\subsection{ Simulation and Discussion}

\section{Implementation of Prototype }


In this section, a mobile application prototype is developed and tested to demonstrate the effectiveness of the proposed system architecture. 

The prototype was developed on a 64-bit Linux Operating System with 6GB RAM and 500GB Disk space. We installed and configured the OpenEMR 4.1.2 
into a web server and  used  the  Eclipse Android development toolkit  to develop the two Android Apps (Doctor and Patient App) which used databases from  the OpenEMR. The implementation challenge was in configuration of the Android application with established databases in OpenEMR.

Three emulators with target application programmer interface (API) 19  were used to test the  short messaging service  \textit {SMS, phone call} and \textit{information retrieval} functions. The functions were enabled by setting its permission in the manifest file of both apps.

The Android Apps retrieves information from the web-server that hosts   OpenEMR database via PhP code (hypertext preprocessor)  files which are given in the web-server. The php files contain MySQL queries that connect to the OpenEMR database and retrieves required information which is passed into a JSON object (a format that is readable by the android application). The information returned from the JSON object is then extracted in the java class by using HashMap ArrayLists and displayed to the android user.

Each PhP file contains MySQL queries for a particular task. The location of the php file is called by the Android App then it executes a line that opens connection to the OpenEMR database. It then closes the connection after executed the query. Below is the PhP line to execute the database connection.  \textit{con = mysql\_connect("localhost","username","password");}

\noindent

where {\it con} is the connection variable, {\it localhost} is the server containing the database, and {\it username} and the {\it password} followed are credentials to the database server. 

\begin{figure}
\centering
\includegraphics[height=7cm]{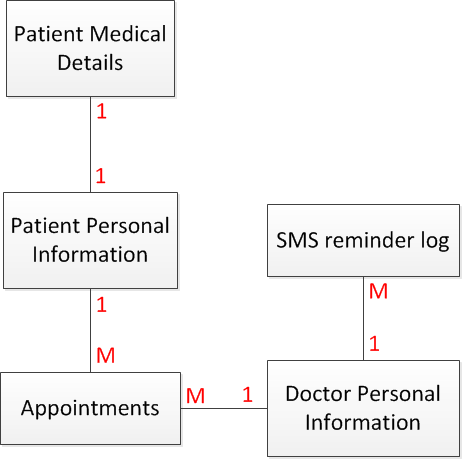}
\caption{Entity relationship diagram of database tables in the proposed system.  }
\label{fig:3}
\end{figure}

The OpenEMR database consists of more than  100 Tables that use MySQL database.   We only used four of them for the scope of this project. We also  added an extra database Table called  ({\it SMS reminder log }) to record the time a medical practitioner sends out SMS appointment reminders. A few minor alterations were also made to a some of the Tables. For instance, we added an extra column for {\it patient password} in order for a patient to log into the app. Also, another column was added in Appointments Table in order to indicate  that the particular appointment has not occurred yet and the value is updated  after appointment has occurred. The relationships (in terms of one-to-one and one-to-many) between the tables utilised is depicted in Figure \ref{fig:3}.

In our medical practitioner prototype, when they click the button to send SMS messages to patients who made appointments for today, it calls a PhP file that searches the {\it Appointments table }for appointments with date set to today's date. The {\it Appointments Table} contains a patient identification number (foreign key) for each appointment record therefore uses that ID to search the {\it Patient Table} in order to retrieve mobile phone number and patient full name which are necessary for generating tailored SMS messages to patients. The data is returned and converted to a JSON object in the java class before it can be displayed to the patient. Figure 4 gives a screen shot of the SMS feature that was sent  to a patient specifying the patient name, appointment time and the medical practitioner. 

\begin{figure}
\centering
\includegraphics[height=7cm]{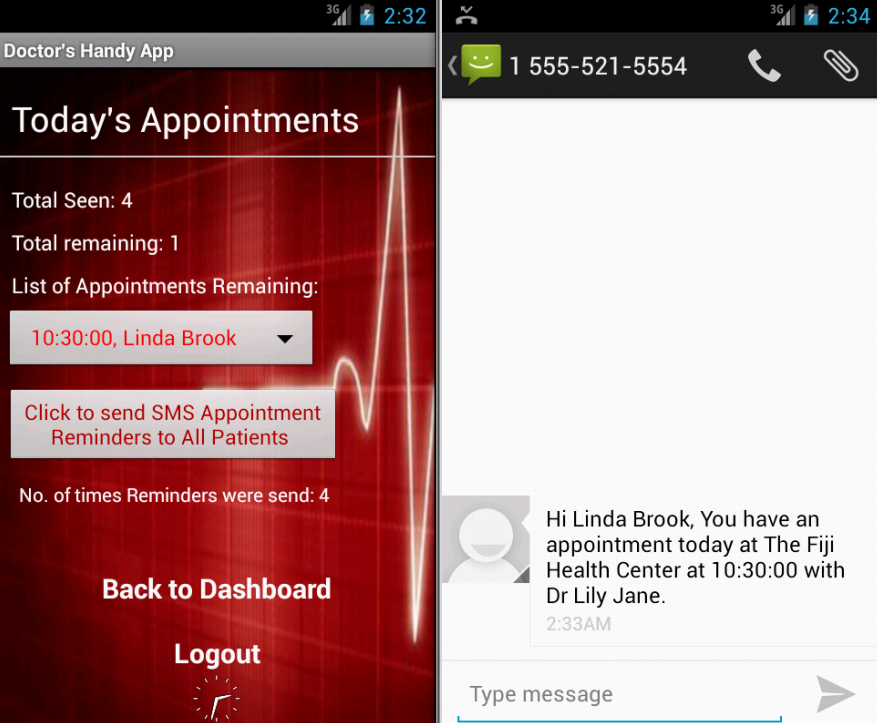}
\caption{Appointment feature in the Medical Practitioner App}
\label{fig:4}
\end{figure}

\begin{figure}
\centering
\includegraphics[height=7cm]{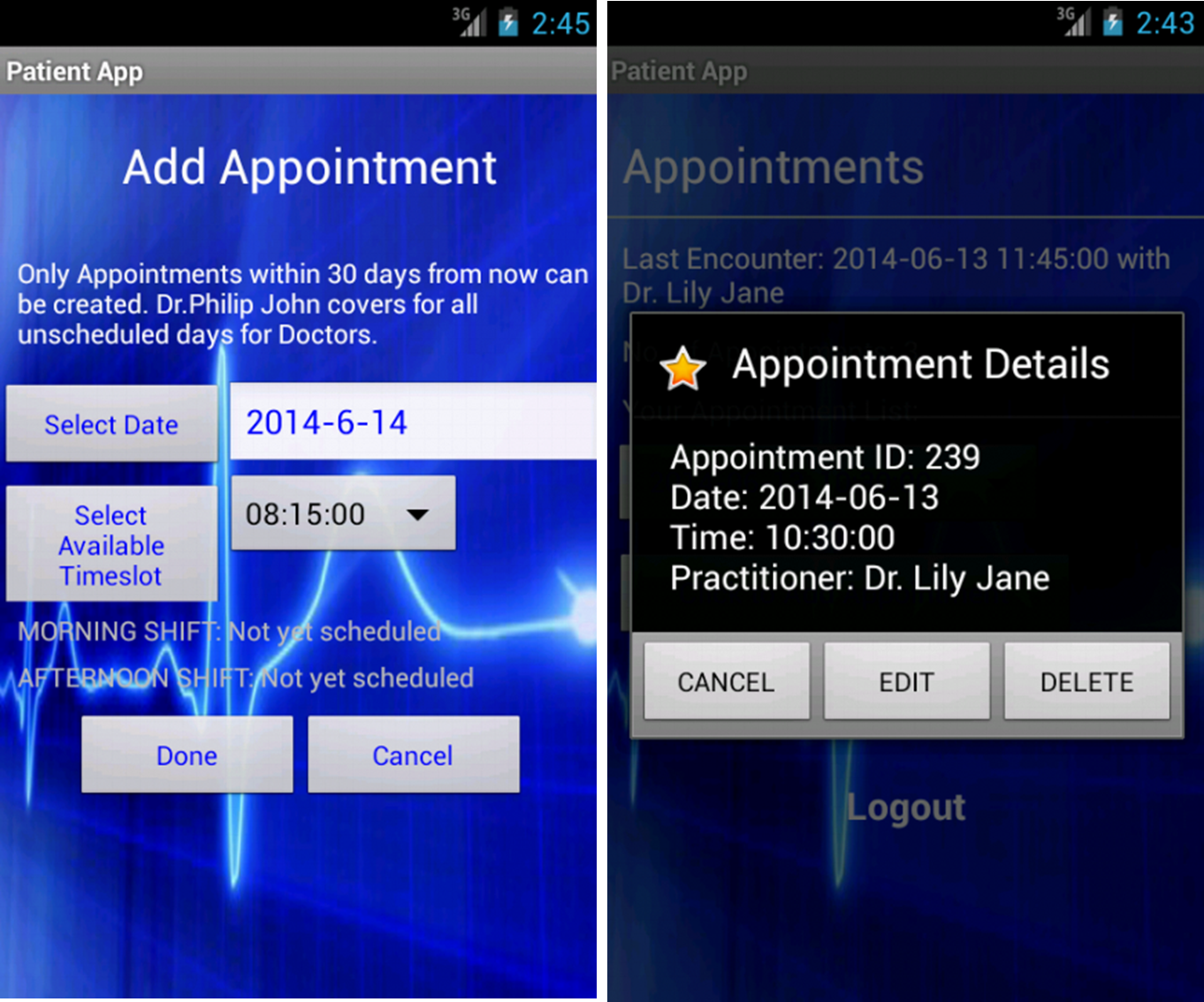}
\caption{Appointment feature in the Patient App}
\label{fig:5}
\end{figure}

The {\it Add Appointment} feature (diagram on left-hand side of Figure 5) allows a patient to select a date from a date picker only from the next day onwards. As shown, the doctor(s) scheduled for the selected day will automatically appear on the {\it Appointment} window thus giving more insight to patients on which doctor will be on duty on which day. If a patient wishes to view his/her appointments, a php file is called to retrieve data from the Appointments table identified by the patient's ID. Figure \ref{fig:5}, the right-hand side diagram captures a scenario where a patient views his/her appointments which are sorted from the nearest appointment to the furthest. If a patient had made at least one appointment, the details of the nearest appointment will always be shown in an alert dialogue and prompting patient to either cancel, edit or remove appointment as seen in Figure  \ref{fig:5} right-hand side diagram.

Managing Appointments using the Patient or Doctor App directly synchronizes with OpenEMR. For the scope of this study, we have only explored the OpenEMR functionalities where the administrator creates doctor schedules, patient appointments (in special cases like appointment cancellation) and enters information for a new patient or doctor. This information can be retrieved from the  Doctor or Patient App. The OpenEMR interface such as that for entering a new appointment is quite different from the one designed for the Patient App therefore separate php files are created for the Patient App.

\begin{figure*}[htb]
\centering
\includegraphics[height=9cm]{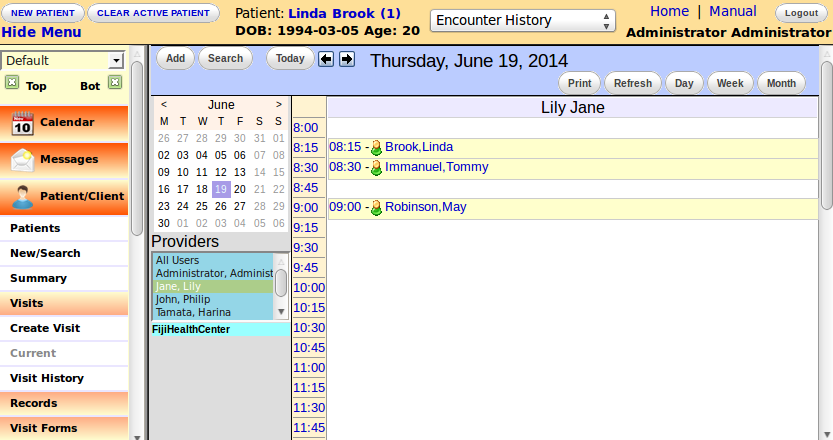}
\caption{Appointment list in OpenEMR electronic medical records system}
\label{fig:6}
\end{figure*}

\subsection{System Evaluation and Discussion}
This section presents an evaluation of the Android applications developed using the mobile application (App) usability check list [33]. We intend to make the apps simple as possible. The issues that we considered for the application interface are outlined in Table 1.\\

\begin{table*}[t]
\begin{tabular}{*{3}{l}}\hline
\centering
\small
{\bf No.} & {\bf Standard} & {\bf How it is applied in the mobile application software (App) }\\\hline
1 & Labels and buttons are clear and concise & Labels give simple descriptions and buttons have simple and well known commands\\\hline
2 & Retains overall consistency and behaviour with mobile platform & The design we use is consistent in terms of user interface -  buttons and background\\\hline
3 & Minimal design & The proposed design is simple and can be easily extended\\\hline
4 & Content is clear and concise & We ensured that instructions were understandable\\\hline
5 & Provides feedback to the user of system status & Doctor App informs doctor if SMS was successfully sent.Patient App lets patient know (via toast message) if appointment was added successfully\\\hline
6 & User interface elements provide visual feedback when pressed & Users are prompted with messages for invalid inputs\\\hline
7 & Colours used provide good contrast and readability & We used dark background behind white texts which is consistent for all activities\\\hline
8 & Font size and spacing ensures good readability & We used a medium-size font\\\hline
9 & Present users with confirmation option when deleting & Delete confirmation is prompted when patient deletes an appointment\\\hline
10 & Speak the users' language (not technical) & The proposed App was in simple English\\\hline
11 & Error messages are free of technical language & Our error messages are simple to understand\\\hline
\end{tabular}
  \caption{Testing software issues}
  \label{tab:table1}
\end{table*}
 
Table \ref{tab:table1} has highlighted that the mobile application prototype is ready to  be deployed in a hospital setting.

Once the system is deployed in a local health centre/hospital in Fiji, then  other tests can also be done as the number of users increases. Server load testing is essential and also a feedback from medical practitioners and patients out the features and quality of the design of the user interface is also needed. Moreover, through the electronic medical record database, it will be possible to generate reports about different type of diseases and their occurrences  according to different weather patterns.  Through the mobile application, it will be easier for medical practitioners to communicate or monitor their patients over time. Since data privacy and security is a major issue, there needs to be a nation wide awareness campaign and special policies will need to be developed. If the policies enable patients to access their records, then the system should have different levels of security and privacy as there are some diseases that need to be confidential for the patients. In that case, the policies can decide if the patient decides what records can be see by the doctor as some information can only be given to doctors about medical history when a certain level of trust is attained. 

The system can also in future be extended to be synchronised with databases and software systems used by  pharmacies so that the exact  drug dispatched  in terms of  brand and quantity  given is recorded. At times, some drugs have certain reactions and if recorded, then reports can be generated and the drug can be monitored. Pharmacists can also enter information whether branded or generic drug was given which can be vital for treatment and it should be recorded if they also cause of certain complications. 

The electronic medical record database can take advantage of cloud computing infrastructure, howsoever, special attention will need to be given at different levels of security. In hospitals and health centres, wireless Internet services would need to be given so that medical practitioners can use the software with use of mobile devices such as  Tablets and smart-phones to update patient information. In emergency departments, such technology can be very useful as treatment of patients without their medical records can be fatal  if medical history is not known.


\section{Conclusion}


We approached the problem of electronic medical records and patient appointments in developing countries using mobile based software system that uses an open source electronic medical record system. 

In general, the proposed system will allow patients to easily access their health records and actively be in contact with medical practitioners. It will improve work productivity and efficiency while reducing cost and waiting times for patients. 

Proper research and planning is needed to identify barriers and to come with strategies for faster adoption of electronic medical record  systems. However, the success of such system implementation will also depend on the support of the medical practitioners  as well as the government policies  and external funding for developing countries. 

It is  essential to design a central data warehouse for patient medical records based on unique national health identification number. The system should  be be able to be used easily by patients and medical practitioners using different mobile devices without limitations to geographical boundaries and cultural - language backgrounds.

Some limitations of the system prototype include user authentication and security of data and would have to be addressed adequately before the proposed system is deployed into the public health care system. Further performance evaluation of the system prototype is necessary in order for large scale implementation.
 
Future work can focus on integrating Android and OpenEMR across other departments of a health care vicinity apart from the outpatient section and also for private practitioners.  In addition, converting the Android applications to  local languages and pacific culture friendly user interface is another future direction.


%

\ifCLASSOPTIONcaptionsoff
  \newpage
\fi


\end{document}